\documentclass[namedreferences]{SolarPhysics}

\usepackage[hyperref,optionalrh]{spr-sola-addons} 
\usepackage{graphicx}        
\usepackage{color}           
\usepackage{breakurl}        

\begin{document}

\begin{article}

\begin{opening}

\title{Frozen-in Fractals All Around: Inferring the Large Scale Effects of Small-Scale Magnetic Structure}

\author{R.T.James~\surname{McAteer}$^{1}$    
       }
\runningauthor{R.T.J. McAteer}
\runningtitle{Frozen-in Fractals}

 \institute{$^{1}$ 	Solar Physics and Space Weather,\\
 		New Mexico State University, \\
 		Las Cruces, NM, USA\\
                     	email: \url{mcateer@nmsu.edu} \\ 
             }

\begin{abstract}
The large-scale structure of the magnetic field in the solar corona provides the energy to power large-scale solar eruptive events. Our physical understanding of this structure, and hence our ability to predict these events, is limited by the type of data currently available. It is shown that the multifractal spectrum is a powerful tool to study this structure, by providing a physical connection between the details of photospheric magnetic gradients and current density at all size scales. This uses concepts associated with geometric measure theory and the theory of weakly differentiable functions to compare Amp\`{e}re's law to the wavelet-transform modulus maximum method. The H\"{o}lder exponent provides a direct measure of the rate of change of current density across spatial size scales. As this measure is independent of many features of the data (pixel resolution, data size, data type, presence of quiet-Sun data), it provides a unique approach to studying magnetic-field complexity and hence a potentially powerful tool for a statistical prediction of solar-flare activity. Three specific predictions are provided to test this theory: the multifractal spectra will not be dependent on the data type or quality; quiet-Sun gradients will not persist with time; structures with large current densities at large size scale will be the source of energy storage for solar eruptive events.
\end{abstract}
\end{opening}

\section{Introduction}
\label{S-Introduction} 
Solar flares and coronal mass ejections occur in active regions in the solar corona; however these magnetic structures are anchored in the solar photosphere. As magnetic flux emerges and is shuffled around the photosphere, the magnetic structure of the coronal field is stressed into non-potential configurations. Energy released in solar eruptive events must be present in the coronal field prior to the onset of the solar flare and/or coronal mass ejection, probably built up over hours to days as the active region either grows (by flux emergence) or as footpoints are shuffled to result in sheared configurations. One of the key challenges in solar information processing is to use data of the photospheric magnetic field to infer the non-potential magnetic energy inherent in the coronal magnetic structure. The precise coronal conditions required to create these energetic events are as yet unknown, as coronal magnetography remains in its infancy. As we lack the ultimate dataset of 3D velocity and magnetic-field information throughout an active region, the prediction of solar flares remains a statistical problem

One approach to this statistical problem is to use the photospheric magnetic-field measurement as the boundary for a 3D extrapolation of coronal structure \citep{der09}. This yields important insights into our knowledge of coronal magnetic fields, one of which is that the final extrapolation is very sensitive to the boundary conditions ({\it i.e.} the spatial resolution, noise levels, image size, and type of magnetic-field data). As the field structure required to store enough energy to power a solar flare must be non-potential, there is often no unique coronal solution. Further the extrapolation is computationally expensive, the required vector data (even in the photosphere) are historically rare, and the chromospheric-field data (where the field may be force-free) remains tantalizingly out of reach. The best datasets in terms of sample size and homogeneity are full disk ({\it i.e.} low spatial resolution) and low-cadence longitudinal field data. The main source of these data are the {\it Global Oscillations Network Group} (GONG) and the {\it Michelson Doppler Imager} (MDI) onboard the {\it Solar and Heliospheric Observatory} (SoHO). Because of this data availability, a lot of research has focused on searching these data for a statistical measure of the coronal free energy from photospheric proxies \citep{sch07, lek07, lek08, blo12} by correlating these proxies with solar flares as observed by GOES. This approach has displayed some potential in terms of a statistical solar-flare prediction, but often these proxies are not well physically motivated and are not well connected to theory. As such, it is difficult to realize the precise conditions under which these proxies fail ({\it i.e.} where they predict a flare that does not occur, and where they fail to predict a flare) and what size scale is most appropriate \citep{ire08}. 

The multifractal spectra may provide a useful proxy of the build-up of magnetic energy for in solar active regions. The proposition often advanced for this connection is that the scaling of magnetic structures is associated with the regime of fully developed turbulence at high Reynolds number. It is also based on the close analytical connection between fractal geometry and turbulence in magnetized plasmas \citep{mca10, mca13}. Empirical studies have used fractals on a large MDI dataset \citep{mca05}, multifractals on a smaller MDI dataset \citep{con08, con10}, the multiscale Fourier index on a small MDI dataset \citep{hew08}, the standard Fourier scaling index on a large dataset \citep{abr05}, and the structure function approach on a small dataset \citep{abr10}. However recent studies by \cite{geo13} and \cite{mca15} reach a contrary conclusion that studies in multifractals do not provide a predictive ability for the onset of solar flares. One potential reason for this discrepancy is that multifractal scaling can also be produced by simple statistical similarity in data of insufficient spatial resolution, and so is not necessarily an indication of increased complexity \citep{mak12}.

This article seeks to close the gap between these contradictory findings by providing a complete analytical connection between multifractal formalism and the set of 3D equations that govern the small-scale and large-scale magnetic structure on the Sun. In Section~\ref{mrn} the background on the appropriate use of multifractals is provided with respect to the frozen-in field structure, arising from the large magnetic Reynolds number in the solar photosphere. Section~\ref{mfs} introduces the concept of multifractals and discusses the wavelet-transform modulus maxima approach of determining this spectrum from 2D images. The multifractal spectra is then connected to the magnetic structure as described by Amp\`{e}re's law in Section~\ref{amp}. Finally Section~\ref{conc} summarizes this approach, discusses why previous results are contradictory, and makes specific predictions of this theory that can be tested empirically in order to make future advances in our understanding of how magnetic energy is stored in the solar corona.

\section{The Magnetic Reynolds Number}
\label{mrn}
It seems that the magnetic-field structure in active regions is the only viable means of storing and releasing sufficient energy to power solar eruptive events. A simple discussion of the equations used to describe this magnetic field leads to a couple of possible choices of characterizing the field. From first principles, the induction equation is given by eliminating the electric field between Faraday's law and Ohm's law
\begin{equation}
\label{e1}
{%
\frac{\partial {\textit {\textbf B}}} {\partial t} = \nabla \times ( {\textit {\textbf v}} \times {\textit {\textbf B}} ) - \nabla \frac{j}{\sigma} ,
}
\end{equation}
which further reduces by Amp\`{e}re's law to
\begin{equation}
\label{e2}
{%
\frac{\partial {\textit {\textbf B}}} {\partial t} = \nabla \times( {\textit {\textbf v}} \times {\textit {\textbf B}} ) - \nabla \times ( \eta\nabla \times {\textit {\textbf B}}) ,
}
\end{equation}
where $\eta  = 1/ \mu_{0} \sigma$ is the magnetic diffusivity. By Gauss' law, and by means of a simple vector identity (and assuming constant $\eta$), this further reduces to the familiar form of the induction equation,
\begin{equation}
\label{e3}
{%
\frac{\partial {\textit {\textbf B}}} {\partial t} = \nabla \times ( {\textit {\textbf v}} \times {\textit {\textbf B}} ) + \eta \nabla^{2} {\textit {\textbf B}} .
}
\end{equation}
This states that any local change in the magnetic field is due to a combination of a convection term and a diffusion term. The battle between these two terms defines whether an active region will grow or shrink with time. The diffusion occurs by buffeting from supergranular flows, and the advection occurs by smaller-scale shuffling (granulation) or flux emergence. The ratio of these two terms is described by the magnetic Reynolds number,
\begin{equation}
\label{e4}
{%
R_{\mathrm{m}}= \frac{\nabla \times ( {\textit {\textbf v}} \times {\textit {\textbf B}} )} { \eta \nabla^{2} {\textit {\textbf B}}} \approx \frac{\upsilon l }{\eta} ,
}
\end{equation}
which acts as an indication of the coupling between the plasma flow field and the magnetic field. For typical photospheric values in solar active regions ($l \approx 10^{5}$ Mm, ${\textit v} \approx 10$ ms$^{-1}$, $\eta \approx 10^{3} $ m$^2$ s$^{-1}$), the magnetic Reynolds number is much greater than one.  In this $R_{\mathrm{m}} \gg 1$ regime, magnetic-field flux is advected with the plasma flow, until such time that gradients are concentrated into short enough length scales that diffusion can balance convection ($l \approx 100$ m), Ohmic dissipation becomes important, and magnetic reconnection can occur. Essentially the large $R_{\mathrm{m}}$ allows for the build-up of energy, followed by a sudden release. Physical system with large $R_{\mathrm{m}}$ are fully turbulent magnetic systems, and this naturally leads to studies of {\em turbulence} and {\em complexity} as potential key parameters to developing a deeper understanding of the physics behind active region evolution and energy storage required for solar eruptive events.

\section{The Multifractal Spectrum}
\label{mfs}
\begin{figure*}
\centering
\includegraphics[width=10cm]{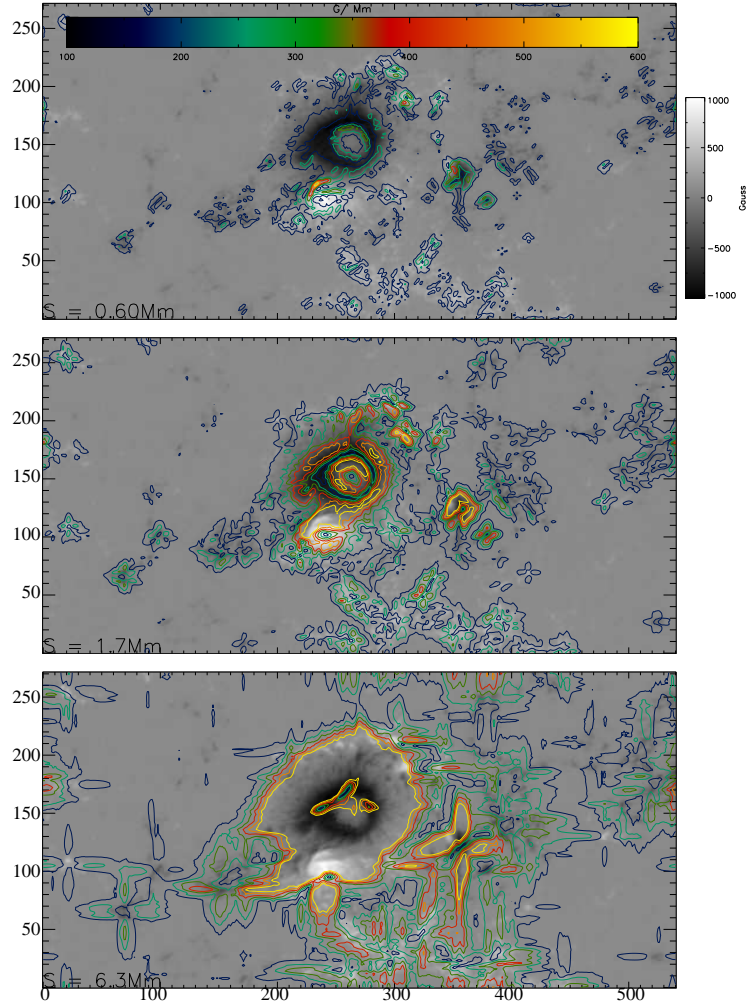}
\caption{A multiscale gradient decomposition of an active region magnetogram. The background image in each row is the same  magnetogram, covering an area of 540Mm by 270Mm, scaled to +/- 1000 G of the vertical color bar. Top: Contours of magnetic gradient (G/Mm) according to the horizontal colorbar, and calculated a spatial scale of 0.6~Mm. Middle: Contours of magnetic gradient on same color scale as top, but calculated at a spatial scale of 1.7~Mm. Bottom: Contours of magnetic gradient on same color scale as top, but calculated at a spatial scale of 6.3~ Mm.}
\label{f1}%
\end{figure*}

The (mono-) fractal dimension of any object is a measure of the self-similarity across all size scales, or the scaling index of any length to area measure, [$A \propto l^{\alpha}$], where $\alpha$ is the capacity dimension, often just called the ``fractal'' dimension. In the more general case, a multifractal system will contain a spectrum of fractal indices of different powers [${f(\alpha)}$] (although strictly this is only true in the asymptotic limit, \citep{Tur08}), and takes account of the fractal measure at each point in space. The $f(\alpha$) curve is equivalent to the $D(h)$ multifractal spectrum, where $D$ is the Haussdorff dimension (or filling dimension) of each H\"{o}lder exponent in the data. If the system only has one H\"{o}lder exponent ({\it i.e.} is a true monofractal) then this $D(h)$ ``spectrum'' will be just one point. If a system contains a large  number of sets of structures with differing H\"{o}lder exponents ({\it i.e.} is a multifractal), then this will be a concave curve. \cite{con10} showed that flare-productive active regions have a multifractal spectrum peaking at large {\it D} ($D>1.2$), agreeing with the monofractal study of \cite{mca05}, and high {\it h} ($h > 0.3$). \cite{kes10} showed that it is possible to use this to separate quiet-Sun gradients from active-region gradients using the wavelet-transform modulus maxima (WTMM). The WTMM method \citep{muz91} provides a novel means to calculate the multifractal spectrum of any dataset. A full description of the method as applied in 1D is available in \cite{mca07} and \cite{mca13b}, as applied to solar flare hard X-ray data, and as applied in 2D by \cite{kes10} and \cite{mca10b}. Briefly, this is achieved by initially computing the 2D continuous wavelet transform [T] of an image [$f(x,y)$]:
\begin{equation}
{\bf T}_\psi [f ] ({\bf b}, a) = \nabla \left( T_\theta [f ] ({\bf b}, a) \right) = \nabla \left( \theta_{b,a} \ast f  \right) \ ,
\label{t_psi}
\end{equation}
at each location [{\bf b}] and where $\psi$ is a smoothing kernel given separably into 
\begin{equation}
\begin{array}{c}
\psi_1 = \partial \theta / \partial x \\ 
\psi_2 = \partial \theta / \partial y . \\ 
\end{array}
\label{mother}
\end{equation}
This amounts to smoothing the data by a Gaussian kernel and applying a gradient filter to the smoothed image to obtain T. The image is then smoothed by a larger Gaussian kernel and the operation is repeated. Figure~\ref{f1} shows gradients at three scales overplotted on magnetic-field data. Note that both quiet-Sun and active-region gradients are detected, as are the ``fake'' gradients at the center of the large negative spot (where the field incorrectly goes to 0 in the center of the largest spot). The largest gradients are at the mixed polarity ``delta'' components at the smallest scale. At the larger scales, the gradients tend to blend together into larger structures over the large-scale polarity-inversion lines evident in the active region. At the largest scale, edge effects creep in as vertical and horizontal streaks. At each scale, edges are extracted as maxima chains where the wavelet-transform modulus [$M_\psi [f ]({\bf b}, a) = | T_\psi [f ]({\bf b}, a)| $] is {\em locally} maximum in the direction of $\mathrm{T}_\psi [f ]({\bf b}, a)$. At each scale, the local maximum is then detected along each chain. A set of maxima lines, [$L_{\bf x0}$] is formed by connecting these points across scale. Statistically, these lines contain all of the information about the local H\"{o}lder exponents [$h$] in the image. Along any one maxima line that connects down to any maxima x0 in the limit at the smallest scale, the wavelet-transform modulus behaves as a power law with exponent $h(x0)$:
\begin{equation}
{%
M_\psi [ f ] [L_{\bf x0} (a)] \sim a^{h({\bf x0})} \ .
}
\label{Hold1}
\end{equation}
By tracking each gradient up through scale, from each location [x0] (where some gradient occurs at the smallest size scale), each of these locations can be characterized by its own H\"{o}lder exponent that prescribes how that gradient varies across scale. The data can then be characterized as a histogram of these $h$-values. A partition-function technique is then used to convert this histogram to the full $D(h)$ multifractal spectrum \citep{kes10} and provide appropriate error bars.

\section{Connecting the Multifractal Spectrum to Magnetic Structure}
\label{amp}
\begin{figure*}
\centering
\includegraphics[width=10cm,angle=90]{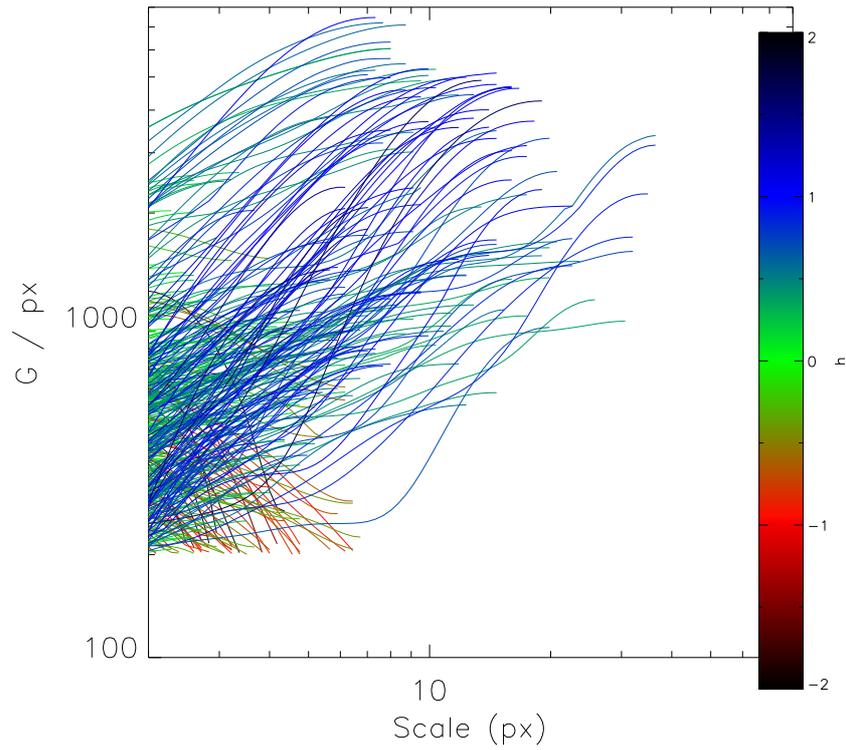}
\caption{The H\"{o}lder exponents of the data presented in Figure~\ref{f1}. Each gradient is tracked to larger and larger scale. The value of the gradient at each size scale is plotted in log-log space. The resulting slope of each line is the H\"{o}lder exponent of each gradient and describes the rate of (spatial) change of current density (Equation~\ref{eqn:Holder}). Red lines show a negative value for $h$, green lines shows values of $h$ around zero, and the blue lines show positive $h$ values where the current density increase for larger scales. The same vertical colorbar is used in Figure~\ref{f3}.}
\label{f2}%
\end{figure*}

The source of magnetic structure in the solar atmosphere is provided by Amp\`{e}re's law,
\begin{equation}
\label{eqn:amp}
{%
\nabla \times {\textit {\textbf B}}  = \mu_{0}{\textit {\textbf J}} .
}
\end{equation}
Where the constant is neglected, this breaks down to a combination of spatial gradients,
\begin{equation}
\label{eqn:curl}
{%
\nabla \times {\textit {\textbf B}} =   \left| \begin{array}{ccc}
i & j & k \\
\partial/\partial x & \partial/\partial y & \partial/\partial z \\
B_x & B_y & B_z \end{array} \right|
=
\begin{array}{l}
\left(\partial B_z / \partial y -  \partial B_y / \partial z \right) {\bf i} + \\
\left(\partial B_x / \partial z -  \partial B_z / \partial x \right) {\bf j} + \\
\left(\partial B_y / \partial x -  \partial B_x / \partial y \right) {\bf k}
\end{array} 
}
\end{equation}
Where we are limited to measurements of $B_z$ ({\it i.e.} longitudinal-field measurements near disk center), we neglect the other spatial gradients involving $B_x$ and $B_y$. This is a good approximation near active regions with strong gradients. Assuming flux [$\int{\textit{\textbf B}}.\mathrm{d}a$] is conserved, then as density falls off exponentially with height, a flux tube will expand in area and so decrease in field strength. Hence $\partial/\partial z$ will fall off rapidly, with the same exponential form as density. Neglecting spatial gradients involving $B_x$ and $B_y$ may not turn out to be a viable assumption in some cases (something to be accounted for in handling full magnetic vector data). Proceeding with $B_z$ only, Equations~(\ref{eqn:amp}) and (\ref{eqn:curl}) become
\begin{equation}
\label{eqn:Bgrad}
{%
{\textit {\textbf J}} (x,y) = \left( \partial B_z / \partial y \right) {\bf i} - \left( \partial B_z / \partial x \right) {\bf j} \ ,
}
\end{equation}
and hence
\begin{equation}
\label{eqn:Bgrad2}
{%
| {\textit {\textbf J}} | = \sqrt{ \left( \partial B_z / \partial y \right)^2 + \left( \partial B_z / \partial x \right)^2} = \nabla_{x,y} B_z \ .
}
\end{equation}
Calculating magnetic spatial gradients is a non-trivial issue. Gradients may occur in any direction, and are not limited to the pixel resolution of the data. The most appropriate way to measure gradients is to do so for every angle and every spatial size scale, and then to study how these vary over size scale, [s], {\it i.e.}
\begin{equation}
\label{eqn:Holder}
{%
\partial | {\textit {\textbf J}} | / \partial s  =  \partial \nabla_{x,y} B_z / \partial s .
}
 \end{equation}
By adopting a derivative of Gaussian as the mother wavelet in the analysis of WTMM in Equations~(\ref{t_psi}) and (\ref{mother}), the local H\"{o}lder exponent, $h({\bf x0})$ in Equation~(\ref{Hold1}) is the rate of change of the gradient of the magnetic field with size scale, {\it i.e.} in the asymptotic limit,
\begin{equation}
h({\bf x0}) = \partial \nabla_{x,y} B_z / \partial s .
\end{equation}
For the gradients identified in Figure~\ref{f1}, the modulus maxima are identified along each ridge, and then tracked up through scale as described in Section~\ref{mfs}. Figure~\ref{f2} shows the result of this tracking procedure. Each gradient is tracked to progressively larger and larger spatial size scales. According to Equation~(\ref{Hold1}) the slope of each of these lines is the H\"{o}lder exponent of each gradient tracked. According to Equation~(\ref {eqn:Holder}), this value is also the spatial rate of change of current density. 

\begin{figure*}
\centering
\includegraphics[width=7cm,angle=90]{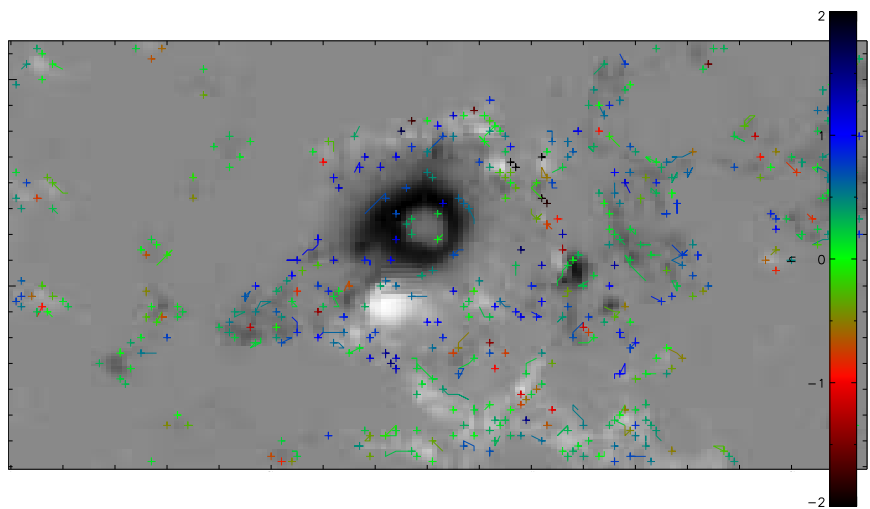}
\caption{The location of each gradient is described by its H\"{o}lder exponent. The negative (red) and small (green) values of gradient are mostly in regions of quiet Sun. The strong positive (blue) H\"{o}lder exponents are the locations where strong gradients exist up to large size scales. These blue gradients delineate the polarity inversion lines within the active region.}
\label{f3}%
\end{figure*}

The lines in Figure~\ref{f2} are color-coded according to the $h$ value. Clearly some gradients exhibit a negative or near-zero $h$. Positive, near-zero H\"{o}lder exponents are ``dust-like'' \citep{geo02}, meaning they are sparse and only exist at small scales ({\it i.e.} the power drops off quickly with increasing scale). By definition in Equation~(\ref{Hold1}), the H\"{o}lder exponent is positive, however recent work \citep{leo14} has extended this analysis into a negative regime via a secondary component. In Equation~(\ref{eqn:Holder}), both negative and near-zero exponents correspond to locations where some current density may be present at small scales, but not at large scales. It is clear from Figure~\ref{f3} that these locations are in the quiet Sun. When tracked over time, these features will not persist and hence they cannot build up the magnetic energy required to be the source of solar flares. 

The positive H\"{o}lder exponents (blue) are large positive $h$. These are ``strong'', meaning they exist at small scales and are larger at larger scales. In Equation~(\ref {eqn:Holder}), these are locations where current density increases at large scale. It is clear from Figure~\ref{f3} that exponents are preferentially located in and around strong magnetic fields. In particular, there are groups of strong $h$ (blue) to the North and South of the active region between the strong negative spot and the positive regions. There is another group of strong $h$ (blue) to the West (right) at the negative polarity plage. All the real active-region gradients present an H\"{o}lder exponent greater than 0.7. The expectation is that when tracked over time, these are the gradients that persist to provide magnetic structure for strong current densities that can exist up to larger spatial scales. It is expected that these are the locations where magnetic energy can therefore build up over time and these will be the locations of energy release.

\begin{figure*}
\centering
\includegraphics[width=8cm,angle=90]{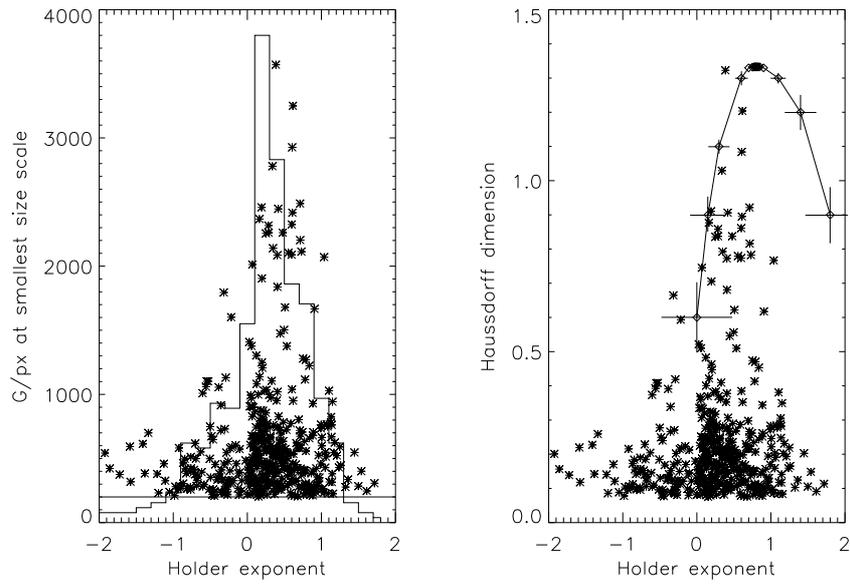}
\caption{(Left: A scatter plot of H\"{o}lder exponent against the gradient at smallest size scale. The histogram of this entire distribution peaks at around $h=0$. Right: The multifractal spectrum is calculated only for those points that have small-scale gradients greater than 200~G/px, horizontal line plotted on the left, and with positive H\"{o}lder exponent. The resulting multifractal spectrum peaks around $D(h=0.8) = 1.35$. Error bars are only plotted for those values where the calculated error is larger than the diamond symbol.}
\label{f4}%
\end{figure*}

Figure~\ref{f4} (left) shows the value of each H\"{o}lder exponent plotted against the gradient at the smallest size scale. Interestingly, a large gradient at small size scale is not necessarily an indicator of a large H\"{o}lder exponent. This shows the need to consider gradients across scale, rather than just gradients on a pixel-by-pixel basis. The frequency histogram of these H\"{o}lder exponents (over plotted) is distinctly bimodal. There is a small negative-$h$ distribution and a larger, long-tailed positive-$h$ distribution. Figure~\ref{f4} (right) shows the same plot of $h$-values, but this time overplotted with the multifractal spectrum preselected only for positive $h$-values that show a large gradient at small size scale, {\it i.e.} this multifractal spectrum is formed from only those data in Figure~\ref{f2} that are blue and show a gradient larger than 200G px$^{-1}$ at the smallest size scale. The multifractal spectrum is calculated using a partition function algorithm applied to the gradients of Figure~\ref{f2} and described in detail by \cite{mca07} and \cite{kes10}. This technique of preselecting H\"{o}lder exponents is justified as the quiet Sun presents its own individual multifractal spectrum; therefore if all the H\"{o}lder exponents are used, the resulting (averaged) multifractal spectrum would be incorrectly broad and would incorrectly peak around $h=0$. By preselecting on both gradient size and H\"{o}lder exponent it is possible to by remove quiet-Sun contamination and therefore extract the active-region multifractal spectra from any dataset.

\section{Conclusions and Future Work}
\label{conc}
The multifractal spectrum exhibits much potential in providing a suitable, and physically motivated, proxy for the complexity (and hence non-potentiality) of active regions. \cite{ire08} showed that a multiscale neutral-line detector could be useful for studying solar flares. Together with the fact that the magnetic Reynold's number is very large, the application of the multifractal spectrum is appropriate. As Amp\`{e}re's law breaks down into a collection of magnetic spatial gradients, the route to obtaining the multifractal spectrum through using these gradients is optimum. By comparing Equation~(\ref{eqn:Bgrad}) for the horizontal components of the current density with Equation~(\ref{Hold1}) that defines the H\"{o}lder exponents  it is shown that the H\"{o}lder exponent explicitly defines the rate of change of current density with spatial size. The histogram of these H\"{o}lder exponents then explicitly defines the overall magnetic complexity of the image. It is important to realize that the histogram in Figure~\ref{f4} is distinctly bimodal. This bimodality also agrees with previous work on solar magnetic fields \citep{kes10, con10} and this bimodality is hard to capture in the usual Legendre-transform approach to the multifractal spectrum. It only becomes evident when each gradient is tracked and studied separately. It can represent a phase transition between two distinct measures. Clearly this bimodality must be studied further across a larger sample of magnetic-field data. This bimodality explains why recent research \citep{geo13} failed to repeat earlier findings of \cite{con08,con10} and \cite{kes10}. If the entire histogram is incorrectly fitted with a single distribution, the resulting incorrect multifractal spectrum will be wide and will peak at lower $D$ and lower $h$. This is a direct result of the influence of the quiet-Sun gradients influencing the overall result. Further, as quiet-Sun gradients are the weakest, it is easy to lose these from image to image, or from instrument to instrument -- hence a cross comparison or study of any time sequence of data that incorrectly includes these quiet-Sun gradients will result in inconsistent, fluctuating, multifractal spectra.
 
It is important to note that the measurement that is taken from the data is that of how the gradient changes with size scale -- this forms the first prediction that the same solution will be determined irrespective of the data type. Future work must test this prediction by comparing the multifractal spectrum on co-temporal and co-spatial regions across datasets from multiple instruments. It is this potential to provide a link between the large MDI dataset, the large GONG dataset, and all future datasets that makes this prediction so intriguing for statistical solar-flare prediction. A second prediction arising from the work presented here is that those gradients defined as quiet Sun - those red and green points in Figure~\ref{f2} and Figure~\ref{f3} - should not persist with time. These gradients can provide some direct heating to the solar atmosphere, but they cannot persist, and hence energy cannot build up to large values of current density. This prediction can be tested by tracking these from image to image -- the gradients defined as quiet Sun will persist for much less time than those gradients defined as active region. A separate study of negative exponents in the quiet Sun may provide insight into these small-scale magnetic flux densities (see discussion by \cite{cha13}). The third prediction is that the energy required to power solar eruptive events comes from those large gradients with large H\"{o}lder exponents. Larger solar flares should result from the largest gradients that persist for the longest time -- indeed it may be possible to calibrate the actual free energy stored in the magnetic field by comparing the size and persistence of both the overall multifractal spectra and the individual gradients with the resulting energy releases. If it can be shown that the devil of the largest scale structures is provided by the detail of the smallest scale gradients, it will then be possible to combine large-scale magnetic proxy studies with smaller-scale extrapolation work to provide future insight into coronal structure, a study that can then be further tested when coronal magnetic-field data become readily available.

\begin{acknowledgment}
This work was supported by National Science Foundation Career award NSS AGS-1255024 and NASA contract NNH12CG10C. The author thanks the referee for useful insight and several very useful references.
\end{acknowledgment}

\section*{Disclosure of Potential Conflicts of Interest}
The author declares he has no conflict of interest



\newpage
\clearpage

\end{article} 

\end{document}